\documentclass{amsart}
\usepackage{amssymb}

\usepackage{graphicx}

\input{tcilatex}
\input tcilatex

\begin{document}
\title[Random walks and weighted graphs]{Random walks pertaining to a class \\
of deterministic weighted graphs}
\author{Thierry Huillet}
\address{Laboratoire de Physique Th\'{e}orique et Mod\'{e}lisation \\
CNRS-UMR 8089 et Universit\'{e} de Cergy-Pontoise \\
2 Avenue Adolphe Chauvin, 95032, Cergy-Pontoise, FRANCE\\
Corresponding author: Thierry.Huillet@u-cergy.fr}
\maketitle

\begin{abstract}
In this note, we try to analyze and clarify the intriguing interplay between
some counting problems related to specific thermalized weighted graphs and
random walks consistent with such graphs.
\end{abstract}

\section{Introduction}

The purpose of this work is to underline the subtle relationship between
some counting problems related to thermalized weighted graphs and random
walks consistent with such graphs; see e.g. \cite{AF} for a detailed
treatise on general random walks on graphs. Let us summarize the topics
developed in this paper. We start with defining finite thermalized weighted
graphs. We show that the notion of a graph singularity spectrum naturally
arises in the problem consisting in counting the number of paths whose
transition-energy rate is asymptotically of a given order. This spectrum is
classically the Legendre transform of the graph pressure function which is
the logarithm of the spectral radius of its weight matrix. The corresponding
Perron eigenvectors play a key role in the Perron-Frobenius theory. We next
recall the Gibbs-variational principle, stating that the pressure produced
by all random walks consistent with the graph structure is bounded below by
the graph pressure which may itself be viewed as the pressure of some
consistent canonical random walk. This turns out to be a by-product of the
Ruelle thermodynamic formalism. We then exhibit and interpret some important
consistent random walks in the light of quasi-stationary distributions for
substochastic random walks. The idea is to normalize the weight matrix of
the graph by its norm to make it substochastic so that, by adding an extra
absorbing coffin state, we may switch to the study of a proper random walk
conditioned to its absorption time. By doing so, a probabilistic
interpretation of both the spectral radius and Perron eigenvectors of the
graph weight matrix naturally comes out and at least two conditioning are
shown to be relevant: one is to condition locally the above random walk on
not hitting the absorbing state in one step at each iteration, the other is
to condition it on not hitting the absorbing state in the remote future. The
latter construction is shown to be the canonical random walk with smallest
pressure production rate. At the end of the paper, we briefly discuss three
particular cases, namely: the case where the weighted graph is reduced to
its adjacency matrix, the case of a potential weighted graph and the case of
a symmetric reversible weighted graph. Several general conclusions that make
use of the above constructions may be drawn. One is an expression in terms
of the average transition energy of the canonical random walk associated to
the adjacency matrix of the value at which the singularity spectrum of any
weighted graph attains its maximum, another is that the entropy production
rate of the locally conditioned random walk\ is always bounded above by the
logarithm of the spectral radius of its adjacency matrix, in the potential
case.

\section{Finite graphs with Boltzmann weights}

Let $W\geq 0$ be some non-negative $N\times N$ weight matrix of some finite
graph (i.e. with non-negative entries $W\left( i,j\right) \geq 0$). Let $%
A=\left[ A\left( i,j\right) \right] $, defined by the indicator function: 
\begin{equation*}
A\left( i,j\right) =\mathbf{I}\left( W\left( i,j\right) >0\right) \in
\left\{ 0,1\right\} ,
\end{equation*}
stand for the Boolean adjacency matrix associated to $W$. With $A^{\prime }$
denoting the transpose of $A$, we shall assume that $A=A^{\prime }$ and that 
$A$ is irreducible: in other words, the underlying topological graph is
undirected and strongly connected so that for each couple of states $\left(
i,j\right) $, there is an integer $m$ such that $A^{m}\left( i,j\right) >0.$
With $\beta \in \Bbb{R}$, we shall choose to represent $W$ under the form: 
\begin{equation}
W\left( i,j\right) =:W_{\beta }\left( i,j\right) =A\left( i,j\right)
e^{-\beta H\left( i,j\right) },  \label{1}
\end{equation}
for some well-behaved transition energies $-\infty <H\left( i,j\right)
<+\infty $ from state $i$ to $j$, not all equal to the same value$.$ The
matrix $W_{\beta }$ therefore appears to be the weight matrix of some
thermalized weighted graph: it can be represented as the Hadamard product
(say, $*$) of $A$ with some positive Boltzmann kernel matrix $K_{\beta }$
with entries $K_{\beta }\left( i,j\right) =e^{-\beta H\left( i,j\right) }$: 
\begin{equation}
W_{\beta }=A*K_{\beta }.  \label{2}
\end{equation}
We note that the Hadamard $\lambda -$power ($\lambda >0$) of $W_{\beta }$
simply is $W_{\beta }^{*\lambda }=W_{\lambda \beta }=A*K_{\lambda \beta }$,
corresponding to a rescaling of $\beta .$\newline

\textbf{Remark:} Let $x_{1}<x_{2}<...<x_{N}$ be $N$ points on the line
(circle). For some matrix $\mathcal{H}$, we may define 
\begin{equation*}
H\left( i,j\right) =\mathcal{H}\left( x_{i},x_{j}\right)
\end{equation*}
to be the interaction energy between sites $\left( i,j\right) $ in positions 
$\left( x_{i},x_{j}\right) $ leading to a slightly more general spatially
extended model that can be treated along similar lines. $\vartriangle $

\subsection{Some counting problems arising in this context}

The quantity 
\begin{equation*}
W_{\beta }\left( \mathbf{i}_{n}\right) :=\prod_{m=1}^{n}W_{\beta }\left(
i_{m-1},i_{m}\right)
\end{equation*}
is the weight of the $n-$path $\mathbf{i}_{n}:=\left\{
i_{0},i_{1},..,i_{n}\right\} $ which is non-null if and only if: $A\left( 
\mathbf{i}_{n}\right) :=\prod_{m=1}^{n}A\left( i_{m-1},i_{m}\right) \neq 0.$
The total product weight of $n-$step paths connecting states $\left(
i_{0},i_{n}\right) $ therefore is given by the corresponding element of the
transfer matrix: 
\begin{equation*}
W_{\beta }^{n}\left( i_{0},i_{n}\right)
=\sum_{i_{1},..,i_{n-1}}\prod_{m=1}^{n}W_{\beta }\left( i_{m-1},i_{m}\right)
=
\end{equation*}
\begin{equation*}
\sum_{i_{1},..,i_{n-1}}\prod_{m=1}^{n}A\left( i_{m-1},i_{m}\right) e^{-\beta
\sum_{m=1}^{n}H\left( i_{m-1},i_{m}\right) }=\sum_{p=1}^{N_{n}\left(
i_{0},i_{n}\right) }e^{-\beta H_{n}\left( p\right) },
\end{equation*}
where $H_{n}\left( p\right) $ is the cumulative energy of the $p-$th path
connecting $\left( i_{0},i_{n}\right) $ and $N_{n}\left( i_{0},i_{n}\right)
:=A^{n}\left( i_{0},i_{n}\right) $ is the number of such $n-$paths. Summing
over the endpoints of the $n-$paths, we obtain the full partition function
of energy 
\begin{equation}
Z_{n}\left( \beta \right) :=\sum_{i_{0},i_{n}}W_{\beta }^{n}\left(
i_{0},i_{n}\right) =\sum_{i_{0},..,i_{n}}\prod_{m=1}^{n}W_{\beta }\left(
i_{m-1},i_{m}\right) =\sum_{p=1}^{N_{n}}e^{-\beta H_{n}\left( p\right) },
\label{3}
\end{equation}
where $N_{n}:=\sum_{i_{0},i_{n}}N_{n}\left( i_{0},i_{n}\right) =Z_{n}\left(
0\right) $ is the total number of $n-$paths. Define

\begin{equation*}
N_{n,\varepsilon }\left( \alpha \right) :=\#\left\{ p\in \left[ N_{n}\right]
:\frac{1}{n}H_{n}\left( p\right) \in \left( \alpha -\varepsilon ,\alpha
+\varepsilon \right) \right\} ,
\end{equation*}
the number of $n-$paths whose transition-energy rate is asymptotically of
order $\alpha .$ We expect 
\begin{equation}
\lim_{\varepsilon \downarrow 0}\lim_{n\uparrow \infty }\frac{1}{n}\log
N_{n,\varepsilon }\left( \alpha \right) =f\left( \alpha \right) \geq 0,
\label{4}
\end{equation}
where $f\left( \alpha \right) =\inf_{\beta }\left( \alpha \beta -p\left(
\beta \right) \right) $, $\alpha \in \left[ \alpha _{-},\alpha _{+}\right] $%
, is the concave Legendre transform of some concave pressure function $%
p\left( \beta \right) $. We may call $f\left( \alpha \right) $ the \emph{%
singularity spectrum of the weighted graph.}

Observing from (\ref{3}) that $Z_{n}\left( \beta \right) =\left\| W_{\beta
}^{n}\right\| $ is a matrix-norm and recalling under our irreducibility
assumption: $\left\| W_{\beta }^{n}\right\| ^{1/n}\rightarrow _{n\uparrow
\infty }\rho _{\beta }$, the spectral radius of $W_{\beta }$, we get 
\begin{equation}
-\frac{1}{n}\log Z_{n}\left( \beta \right) \underset{n\uparrow \infty }{%
\rightarrow }p\left( \beta \right) =-\log \rho _{\beta }.  \label{5}
\end{equation}
For each $\beta $, the number $\rho _{\beta },$ as an eigenvalue of $%
W_{\beta }$, satisfies det$\left( \frac{1}{\rho _{\beta }}W_{\beta
}-I\right) =0.$

For each $n$, the quantity 
\begin{equation*}
\Phi _{n}\left( \beta \right) :=\frac{Z_{n}\left( \beta \right) }{%
Z_{n}\left( 0\right) }=\frac{1}{N_{n}}\sum_{p=1}^{N_{n}}e^{-\beta
H_{n}\left( p\right) }
\end{equation*}
is the Laplace-Stieltjes transform of some discrete probability measure on $%
n-$paths satisfying $-\frac{1}{n}\log \Phi _{n}\left( \beta \right)
\rightarrow p\left( \beta \right) -p\left( 0\right) =-\log \frac{\rho
_{\beta }}{\rho _{0}}$. This limit therefore is the log-Laplace transform of
some probability distribution which, in particular, is smooth and concave.
The pressure $p\left( \beta \right) $ is classically related to the scaled
free energy $\tau \left( \beta \right) $ by: $p\left( \beta \right) =:\beta
\tau \left( \beta \right) .$ Note that, for each $n$, we also have: 
\begin{equation*}
\Phi _{n}\left( \beta \right) =\frac{1}{N_{n}}\sum_{h_{n}\in \mathcal{H}%
_{n}}N_{n}\left( h_{n}\right) e^{-\beta h_{n}},
\end{equation*}
where $\mathcal{H}_{n}:=$ Span$\left( H_{n}\left( p\right) ;p\in \left[
N_{n}\right] \right) ,$ $N_{n}\left( h_{n}\right) :=\#\left\{ p\in \left[
N_{n}\right] :H_{n}\left( p\right) =h_{n}\right\} $ and $N_{n}\left(
h_{n}\right) /N_{n}$ is the probability of $n-$paths of energy $h_{n}.$ 
\newline

With $\mathbf{1}=\left( 1,1,..,1\right) ^{\prime }$, we define $\mathbf{w}%
_{\beta }:=W_{\beta }\mathbf{1}$ to be the column-sum vector of $W_{\beta }$%
, with entries $w_{\beta }\left( i\right) =\sum_{j}A\left( i,j\right)
e^{-\beta H\left( i,j\right) }.$ Define $w_{\beta }^{+}=\max_{i}w_{\beta
}\left( i\right) $, $w_{\beta }^{-}=\min_{i}w_{\beta }\left( i\right) $. We
have $w_{\beta }^{-}\leq \rho _{\beta }\leq w_{\beta }^{+}$ and so 
\begin{equation*}
-\log w_{\beta }^{+}\leq p\left( \beta \right) \leq -\log w_{\beta }^{-}.
\end{equation*}
Clearly, it holds that\newline

$-\log w_{\beta }^{+}\underset{\beta \uparrow \infty }{\sim }\beta \alpha
_{-}$ where $\alpha _{-}=\min_{i,j:A\left( i,j\right) =1}H\left( i,j\right) $
and similarly,

$-\log w_{\beta }^{-}\underset{\beta \downarrow -\infty }{\sim }\beta \alpha
_{+}$ where $\alpha _{+}=\max_{i,j:A\left( i,j\right) =1}H\left( i,j\right)
>\alpha _{-}$.\newline

We can check that $f\left( \alpha _{-}\right) =f\left( \alpha _{+}\right) =0$
and that the maximum of $f\left( \alpha \right) $ is attained at $\alpha
=\alpha _{0}=p^{\prime }\left( 0\right) $. We also have $f\left( \alpha
\right) =$ $f\left( p^{\prime }\left( \beta \right) \right) =p^{\prime
}\left( \beta \right) \beta -p\left( \beta \right) $ so that $f\left( \alpha
_{0}\right) =-p\left( 0\right) =\log \rho _{0}>0$ where $\rho _{0}$ is the
spectral radius of $A.$\newline

Whenever $\alpha _{-}<0$, $\alpha _{+}>0$, then there is a $\beta _{c},$
possibly not equal to $0,$ given by: $p^{\prime }\left( \beta _{c}\right)
=0. $ With $\alpha _{c}:=p^{\prime }\left( \beta _{c}\right) =0$, $f\left(
\alpha _{c}\right) =-p\left( \beta _{c}\right) >0.$\newline

For all distinct pairs of nodes $\left( i,j\right) $, let $m\left(
i,j\right) =\inf \left( m>1:A^{m}\left( i,j\right) >0\right) .$ Then 
\begin{equation*}
m_{*}=\max_{\left( i,j\right) }m\left( i,j\right)
\end{equation*}
is the diameter of the adjacency graph. For each $\left( i,j\right) $, there
can be more than one path of minimal length $m\left( i,j\right) .$ Let $%
N_{m\left( i,j\right) }\geq 1$ be the number of such length$-m\left(
i,j\right) $ paths and let $h\left( i,j\right) $ be the energy of any path
with smallest energy among these $N_{m\left( i,j\right) }$ paths. Then 
\begin{equation*}
\alpha _{*}=\max_{\left( i,j\right) }\left[ h\left( i,j\right) /m\left(
i,j\right) \right]
\end{equation*}
is a quantity of interest related to the energy diameter of the weighted
graph. Clearly, $\alpha _{-}<\alpha _{*}<$ $\alpha _{+}$ and $\alpha _{*}$
belongs to the range of the spectrum.

\subsection{Perron-Frobenius and the like}

Let $\mathbf{\pi }_{\beta }^{\prime }>0$ and $\mathbf{\varphi }_{\beta }%
\mathbf{>0}$ be the line and column ($l_{1}-$norm $1$) Perron vectors of $%
W_{\beta }$ associated to the spectral radius $\rho _{\beta }$ of $W_{\beta
} $: 
\begin{equation}
\rho _{\beta }\mathbf{\pi }_{\beta }^{\prime }=\mathbf{\pi }_{\beta
}^{\prime }W_{\beta }\text{ and }\rho _{\beta }\mathbf{\varphi }_{\beta
}=W_{\beta }\mathbf{\varphi }_{\beta }\mathbf{.}  \label{6}
\end{equation}
Under our hypothesis, $\rho _{\beta }>0$ is the algebraically simple real
dominant eigenvalue of $W_{\beta }$. If $A$ is in addition primitive ($%
A^{m}>0$ for some integer $m$), then all the other eigenvalues of $W_{\beta
} $ are strictly contained within the disk: $\left| \rho \right| <\rho
_{\beta }$, else some could lie on the disk $\left| \rho \right| =\rho
_{\beta }$ because of the underlying periodicity of the problem.

We shall let $\mathbf{\phi }_{\beta }\mathbf{:=\varphi }_{\beta }\mathbf{/}%
\left( \mathbf{\pi }_{\beta }^{\prime }\mathbf{\varphi }_{\beta }\right) $
in such a way that the Hadamard product of $\mathbf{\pi }_{\beta }$ and $%
\mathbf{\phi }_{\beta }$, namely the column vector $\mathbf{\pi }_{\beta }*%
\mathbf{\phi }_{\beta }$, with components $\mathbf{\pi }_{\beta }*\mathbf{%
\phi }_{\beta }\left( i\right) =\pi _{\beta }\left( i\right) \phi _{\beta
}\left( i\right) $, has $l_{1}-$norm $1 $ (i.e. $\mathbf{\pi }_{\beta
}^{\prime }\mathbf{\phi }_{\beta }=1$).\newline

\textbf{Remarks:} $\left( i\right) $ When $H\left( i,j\right) =H\left(
j,i\right) $, $W_{\beta }$ is itself symmetric ($W_{\beta }=W_{\beta
}^{\prime }$) then $\mathbf{\pi }_{\beta }=\mathbf{\varphi }_{\beta }$ and $%
\mathbf{\pi }_{\beta }*\mathbf{\phi }_{\beta }=\mathbf{\varphi }_{\beta }*%
\mathbf{\varphi }_{\beta }\mathbf{/}\left( \mathbf{\varphi }_{\beta
}^{\prime }\mathbf{\varphi }_{\beta }\right) $. Then it is useful to
introduce the \emph{probability wave} vector of $l_{2}-$norm $1:$ $\mathbf{%
\psi }_{\beta }=\mathbf{\varphi }_{\beta }\mathbf{/}\left( \mathbf{\varphi }%
_{\beta }^{\prime }\mathbf{\varphi }_{\beta }\right) ^{1/2}$ in such a way
that $\mathbf{\pi }_{\beta }*\mathbf{\phi }_{\beta }=\mathbf{\psi }_{\beta }*%
\mathbf{\psi }_{\beta }$.\newline

$\left( ii\right) $ Letting $\varepsilon _{\beta }:=1-\rho _{\beta
}/w_{\beta }^{+}\geq 0$ stand for the scaled spectral gap of the graph, the
equation giving the right eigenvector $\mathbf{\varphi }_{\beta }$ may be
recast as 
\begin{equation*}
\varepsilon _{\beta }\mathbf{\varphi }_{\beta }=\left( I-\frac{1}{w_{\beta
}^{+}}W_{\beta }\right) \mathbf{\varphi }_{\beta }=:-\Delta _{\beta }\mathbf{%
\varphi }_{\beta },
\end{equation*}
where $\Delta _{\beta }=\frac{1}{w_{\beta }^{+}}W_{\beta }-I$ is a Laplacian
of the graph. Observe that $w_{\beta }^{+}=\left\| \left| W_{\beta }\right|
\right\| _{\infty }$ is the matrix norm induced by the $l_{\infty }-$vector
norm so that $\rho _{\beta }/w_{\beta }^{+}\leq 1$ (i.e. $\varepsilon
_{\beta }\geq 0$) and that $\log \rho _{\beta }/\log w_{\beta
}^{+}\rightarrow _{\beta \uparrow \infty }1.$ $\triangledown $\newline

Consistently with (\ref{6}), we shall let $\mathbf{\pi }_{0}^{\prime }>0$
and $\mathbf{\varphi }_{0}\mathbf{>0}$ stand for the line and column Perron
vectors of $A=W_{0}$ with: 
\begin{equation*}
\rho _{0}\mathbf{\pi }_{0}^{\prime }=\mathbf{\pi }_{0}^{\prime }A\text{ and }%
\rho _{0}\mathbf{\varphi }_{0}=A\mathbf{\varphi }_{0}\mathbf{,}
\end{equation*}
associated to the spectral radius $\rho _{0}$ of $A=W_{0}.$ We shall let $%
\mathbf{\phi }_{0}\mathbf{:=\varphi }_{0}\mathbf{/}\left( \mathbf{\pi }%
_{0}^{\prime }\mathbf{\varphi }_{0}\right) $ so that the Hadamard product $%
\mathbf{\pi }_{0}*\mathbf{\phi }_{0}$ has $l_{1}-$norm $1.$ Note that, since 
$A=A^{\prime }$, $\mathbf{\pi }_{0}=\mathbf{\varphi }_{0}$ and $\mathbf{\pi }%
_{0}*\mathbf{\phi }_{0}=\mathbf{\varphi }_{0}*\mathbf{\varphi }_{0}\mathbf{/}%
\left( \mathbf{\varphi }_{0}^{\prime }\mathbf{\varphi }_{0}\right) =:\mathbf{%
\psi }_{0}*\mathbf{\psi }_{0}$ where $\mathbf{\psi }_{0}=\mathbf{\varphi }%
_{0}\mathbf{/}\left( \mathbf{\varphi }_{0}^{\prime }\mathbf{\varphi }%
_{0}\right) ^{1/2}$ is the wave vector associated to $A.$

\section{Random walks on graphs}

In the study of weighted graphs, questions pertaining to counting are then
relevant. Once such weighted graphs have been introduced, it is useful to
consider the following particular class of random walks attached to such
graphs.

Let $0\leq \Pi =\left[ \Pi \left( i,j\right) \right] $ denote some
stochastic matrix with column sums one: $\Pi \mathbf{1}=\mathbf{1}.$ Let $%
\mathcal{P}_{W_{\beta }}$ be the set of stochastic matrices which are $%
W_{\beta }-$consistent in the sense that 
\begin{equation}
\Pi \in \mathcal{P}_{W_{\beta }}\Leftrightarrow \left\{ \Pi \left(
i,j\right) =0\text{ whenever }W_{\beta }\left( i,j\right) =0\right\} .
\label{7}
\end{equation}
Let $\mathbf{\mu }^{\prime }>0$ be the line left Perron eigenvector of $\Pi $%
, satisfying$:$ $\mathbf{\mu }^{\prime }=\mathbf{\mu }^{\prime }\Pi $ (the
unique invariant probability measure associated to $\Pi $). Clearly, to each
such $\Pi $ a positive recurrent random walk (RW, say) can be associated.

\subsection{A variational principle and first consequences}

In this context, the following Gibbs-variational principle indeed holds \cite
{AGD}, resulting from the Ruelle thermodynamic formalism \cite{R}. It reads: 
\begin{equation*}
\log \rho _{\beta }=\sup_{\Pi \in \mathcal{P}_{W_{\beta }}}\left(
-\sum_{i}\mu \left( i\right) \sum_{j}\Pi \left( i,j\right) \log \Pi \left(
i,j\right) +\sum_{i}\mu \left( i\right) \sum_{j}\Pi \left( i,j\right) \log
W_{\beta }\left( i,j\right) \right)
\end{equation*}
\begin{equation}
=\sup_{\Pi \in \mathcal{P}_{W_{\beta }}}\left( -\sum_{i}\mu \left( i\right)
\sum_{j}\Pi \left( i,j\right) \log \Pi \left( i,j\right) -\beta \sum_{i}\mu
\left( i\right) \sum_{j}\Pi \left( i,j\right) H\left( i,j\right) \right) ,
\label{8}
\end{equation}
where the supremum is attained for the unique stochastic matrix $\Pi _{*}$
which is $W_{\beta }-$consistent and defined by the Doob transform: 
\begin{equation}
\Pi _{*}\left( i,j\right) =\frac{1}{\rho _{\beta }}W_{\beta }\left(
i,j\right) \frac{\phi _{\beta }\left( j\right) }{\phi _{\beta }\left(
i\right) }.  \label{9}
\end{equation}
With $D_{\mathbf{\phi }_{\beta }}:=$ diag$\left( \mathbf{\phi }_{\beta
}\right) $, this is also; $\Pi _{*}=\frac{1}{\rho _{\beta }}D_{\mathbf{\phi }%
_{\beta }}^{-1}W_{\beta }D_{\mathbf{\phi }_{\beta }}$, in matrix form. The
corresponding invariant measure satisfying $\mathbf{\mu }_{*}^{\prime }=%
\mathbf{\mu }_{*}^{\prime }\Pi _{*}$ can easily be checked to be: 
\begin{equation}
\mathbf{\mu }_{*}=\mathbf{\pi }_{\beta }*\mathbf{\phi }_{\beta },  \label{10}
\end{equation}
the Hadamard product of the left and right eigenvectors of the weight matrix 
$W_{\beta }$. We shall call the RW with transition probability $\Pi _{*}$
the \emph{canonical} RW consistent with $W_{\beta }$. Using this canonical
RW construction, we get: 
\begin{equation*}
\log \rho _{\beta }=-\frac{1}{\rho _{\beta }}\sum_{i}\pi _{\beta }\left(
i\right) \phi _{\beta }\left( i\right) \sum_{j}W_{\beta }\left( i,j\right) 
\frac{\phi _{\beta }\left( j\right) }{\phi _{\beta }\left( i\right) }\log
\left( \frac{1}{\rho _{\beta }}\frac{\phi _{\beta }\left( j\right) }{\phi
_{\beta }\left( i\right) }\right)
\end{equation*}
\begin{equation*}
=-\frac{1}{\rho _{\beta }}\sum_{i}\pi _{\beta }\left( i\right)
\sum_{j}W_{\beta }\left( i,j\right) \phi _{\beta }\left( j\right) \left[
-\log \rho _{\beta }+\log \phi _{\beta }\left( j\right) -\log \phi _{\beta
}\left( i\right) \right]
\end{equation*}
\begin{equation*}
=\log \rho _{\beta }+\sum_{i}\pi _{\beta }\left( i\right) \phi _{\beta
}\left( i\right) \log \phi _{\beta }\left( i\right) -\frac{1}{\rho _{\beta }}%
\sum_{i}\pi _{\beta }\left( i\right) \sum_{j}W_{\beta }\left( i,j\right)
\phi _{\beta }\left( j\right) \log \phi _{\beta }\left( j\right) ,
\end{equation*}
leading to the expression 
\begin{equation*}
\rho _{\beta }=\frac{\sum_{i}\pi _{\beta }\left( i\right) \sum_{j}W_{\beta
}\left( i,j\right) \phi _{\beta }\left( j\right) \log \phi _{\beta }\left(
j\right) }{\sum_{i}\pi _{\beta }\left( i\right) \phi _{\beta }\left(
i\right) \log \phi _{\beta }\left( i\right) },
\end{equation*}
in terms of the left and right Perron eigenvectors of $W_{\beta }.$ As a
result, we obtain 
\begin{equation}
p\left( \beta \right) =:\beta \tau \left( \beta \right) =-\log \left[ \frac{%
\sum_{i}\pi _{\beta }\left( i\right) \sum_{j}A\left( i,j\right) e^{-\beta
H\left( i,j\right) }\phi _{\beta }\left( j\right) \log \phi _{\beta }\left(
j\right) }{\sum_{i}\pi _{\beta }\left( i\right) \phi _{\beta }\left(
i\right) \log \phi _{\beta }\left( i\right) }\right] .  \label{11}
\end{equation}

From (\ref{8}), for all $W_{\beta }-$consistent stochastic matrix $\Pi \neq
\Pi _{*}:$%
\begin{equation}
\log \rho _{\beta }>-\sum_{i}\mu \left( i\right) \sum_{j}\Pi \left(
i,j\right) \log \Pi \left( i,j\right) -\beta \sum_{i}\mu \left( i\right)
\sum_{j}\Pi \left( i,j\right) H\left( i,j\right) .  \label{11a}
\end{equation}
In the right-hand-side of (\ref{11a}), $s:=-\sum_{i}\mu \left( i\right)
\sum_{j}\Pi \left( i,j\right) \log \Pi \left( i,j\right) =:\sum_{i}\mu
\left( i\right) s\left( i\right) $ is the equilibrium Shannon entropy
production rate of the ergodic Markov chain governed by $\Pi $ and $%
u:=\sum_{i}\mu \left( i\right) \sum_{j}\Pi \left( i,j\right) H\left(
i,j\right) =:\sum_{i}\mu \left( i\right) u\left( i\right) $ its equilibrium
internal transition energy. It follows from (\ref{8}) that the quantity $%
p\left( \beta \right) :=-\log \rho _{\beta }$ is a universal lower bound for
the equilibrium pressure production rate of all $W_{\beta }-$consistent
walkers. Stated differently, defining the pressure of a consistent RW
governed by $\Pi \in \mathcal{P}_{W_{\beta }}$, $\Pi \neq \Pi _{*}$, as: 
\begin{equation*}
p_{\Pi }\left( \beta \right) :=\beta \sum_{i}\mu \left( i\right) \sum_{j}\Pi
\left( i,j\right) H\left( i,j\right) +\sum_{i}\mu \left( i\right)
\sum_{j}\Pi \left( i,j\right) \log \Pi \left( i,j\right) ,
\end{equation*}
it holds that: 
\begin{equation}
\text{for all }\Pi \neq \Pi _{*}\in \mathcal{P}_{W_{\beta }},\text{ }p_{\Pi
}\left( \beta \right) >p\left( \beta \right) =p_{\Pi _{*}}\left( \beta
\right) .  \label{12}
\end{equation}
\newline

\textbf{Remark:} On the other hand, we also recall the Friedland-Karlin
inequality \cite{FK} of a similar flavor: 
\begin{equation}
p\left( \beta \right) =-\log \rho _{\beta }\geq \sum_{i}\pi _{\beta }\left(
i\right) \phi _{\beta }\left( i\right) \log w_{\beta }\left( i\right)
\label{12a}
\end{equation}
where, $\mathbf{w}_{\beta }:=W_{\beta }\mathbf{1}$ is the column-sum vector
of $W_{\beta }:$ $w_{\beta }\left( i\right) =\sum_{j}A\left( i,j\right)
e^{-\beta H\left( i,j\right) }.$ It gives a universal lower bound of $%
p\left( \beta \right) $ in terms of the invariant measure $\mu _{*}\left(
i\right) =\pi _{\beta }\left( i\right) \phi _{\beta }\left( i\right) $
associated to $\Pi _{*}$. $\vartriangle $

\subsection{\textbf{The entropy production rate of the RW governed by }$\Pi
\in \mathcal{P}_{W_{\beta }}$}

We need to say a few words on the way to compute the quantity $s$ associated
to some $\Pi $. We refer to \cite{RAC} for additional information. For each
pair of connecting states $\left( i_{0},i_{n}\right) ,$ define: 
\begin{equation*}
\Pi _{\lambda }^{n}\left( i_{0},i_{n}\right)
:=\sum_{i_{1},..,i_{n-1}}\prod_{m=1}^{n}\Pi \left( i_{m-1},i_{m}\right)
^{\lambda },
\end{equation*}
where $\Pi \left( i_{m-1},i_{m}\right) ^{\lambda }$ is the $\left(
i_{m-1},i_{m}\right) $ entry of $\Pi ^{*\lambda },$ the Hadamard $\lambda -$%
power of $\Pi $, $\lambda >0.$ Define the R\'{e}nyi $\lambda -$entropy of
all $n-$paths of the RW governed by $\Pi $ and started using the invariant
measure $\mathbf{\mu }$ to be 
\begin{equation*}
R_{n}\left( \lambda \right) :=\frac{1}{1-\lambda }\log \sum_{i_{0},i_{n}}\mu
\left( i_{0}\right) \Pi _{\lambda }^{n}\left( i_{0},i_{n}\right) .
\end{equation*}
Note that with $\Pi \left( \mathbf{i}_{n}\right) :=\prod_{m=1}^{n}\Pi \left(
i_{m-1},i_{m}\right) $ the probability of the $n-$path $\mathbf{i}%
_{n}:=\left\{ i_{0},i_{1},..,i_{n}\right\} :$%
\begin{equation*}
R_{n}\left( \lambda \right) \rightarrow _{\lambda \uparrow 1}S_{n}=-\sum_{%
\mathbf{i}_{n}}\mu \left( i_{0}\right) \Pi \left( \mathbf{i}_{n}\right) \log
\Pi \left( \mathbf{i}_{n}\right) ,
\end{equation*}
the Shannon entropy of $n-$paths at equilibrium. Then, with $\alpha \left(
\lambda \right) :=\sum_{i}\mu \left( i\right) \sum_{j}\Pi \left( i,j\right)
^{\lambda }:$%
\begin{equation}
\frac{1}{n}R_{n}\left( \lambda \right) \underset{n\uparrow \infty }{%
\rightarrow }r\left( \lambda \right) :=\frac{1}{1-\lambda }\log \alpha
\left( \lambda \right)  \label{13}
\end{equation}
where $r\left( \lambda \right) $ is the R\'{e}nyi-entropy production rate of
the walker. As a result, 
\begin{equation*}
r\left( \lambda \right) \rightarrow _{\lambda \uparrow 1}s=-\sum_{i}\mu
\left( i\right) \sum_{j}\Pi \left( i,j\right) \log \Pi \left( i,j\right)
=-\alpha ^{\prime }\left( 1\right) .
\end{equation*}
This approach is useful to compute the Shannon-entropy production rate $s$
for specific $\Pi $s.

\subsection{Random walks consistent with $W_{\beta }$}

We now exhibit and interpret some important $W_{\beta }-$consistent RWs in
the light of \emph{quasi-stationary distributions} for substochastic RWs. By
doing so, a probabilistic interpretation of $\rho _{\beta },$\emph{\ }$%
\mathbf{\pi }_{\beta }$\emph{\ }and\emph{\ }$\mathbf{\phi }_{\beta }$
emerges.

Let us first normalize $W_{\beta }$ in the following way. Consider the
matrix: 
\begin{equation}
\overline{W}_{\beta }:=\frac{W_{\beta }}{\left\| W_{\beta }\right\| },
\label{14}
\end{equation}
for some matrix-norm $\left\| W_{\beta }\right\| $ of $W_{\beta }.$ For
example: $\left\| W_{\beta }\right\| =N\max_{i,j}W_{\beta }\left( i,j\right) 
$ or $\left\| W_{\beta }\right\| =\sum_{i,j}W_{\beta }\left( i,j\right) $ or 
$\left\| W_{\beta }\right\| =w_{\beta }^{+}=\max_{i}w_{\beta }\left(
i\right) $.

The spectral radius of $\overline{W}_{\beta }$ now is $\overline{\rho }%
_{\beta }=\rho _{\beta }/\left\| W_{\beta }\right\| <1$ with the same left
and right strictly positive Perron eigenvectors $\mathbf{\pi }_{\beta }>0$
and $\mathbf{\varphi }_{\beta }\mathbf{>0}$ as for $W_{\beta }$ in (\ref{6})$%
.$ By doing so, the matrix $\overline{W}_{\beta }$ is \emph{substochastic}
in the sense that, with $\overline{\mathbf{w}}_{\beta }:=\overline{W}_{\beta
}\mathbf{1}$ the column-sum vector of $\overline{W}_{\beta }$, then: $%
\overline{w}_{\beta }\left( i\right) \in \left( 0,1\right] $ with $\overline{%
w}_{\beta }\left( i\right) <1$ for at least one $i$. To recast this problem
into a stochastic problem, we may add an additional \emph{coffin state}, say 
$\partial :=\left\{ 0\right\} $ and look at the enlarged $\left( N+1\right)
\times \left( N+1\right) $ stochastic matrix $P:$%
\begin{equation}
P=\left[ 
\begin{array}{ll}
1 & \mathbf{0}^{\prime } \\ 
\mathbf{1-}\overline{\mathbf{w}}_{\beta } & \overline{W}_{\beta }
\end{array}
\right] .  \label{15}
\end{equation}
$P$ now is the stochastic transition matrix of a RW, say $\left\{
X_{n}\right\} ,$ having state $\left\{ 0\right\} $ as an additional \emph{%
absorbing} state. Let $\tau _{0}$ be the first hitting time of $\partial
=\left\{ 0\right\} $ for this RW $\left\{ X_{n}\right\} $.

Using this construction, clearly, the substochastic matrix $\overline{W}%
_{\beta }$ turns out to be the transition matrix of the process $X_{n}\cdot 
\mathbf{I}\left( \tau _{0}>n\right) $ (i.e. $X_{n},$ restricted to the set $%
\tau _{0}>n$). In other words, with $\mathbf{e}_{i_{0}}^{\prime }$ the
line-vector with a single $1$ in position $i_{0}$, $0$ elsewhere, we have 
\begin{equation*}
\Bbb{P}_{i_{0}}\left( X_{n}=i_{n},\tau _{0}>n\right) =\overline{W}_{\beta
}^{n}\left( i_{0},i_{n}\right) =\mathbf{e}_{i_{0}}^{\prime }\overline{W}%
_{\beta }^{n}\mathbf{e}_{i_{n}},\text{ }i_{0},i_{n}\in \left\{ 1,.,N\right\}
.
\end{equation*}
Therefore, 
\begin{equation}
\Bbb{P}_{i_{0}}\left( \tau _{0}>n\right) =\mathbf{e}_{i_{0}}^{\prime }%
\overline{W}_{\beta }^{n}\mathbf{1}.  \label{15a}
\end{equation}
We note that $\Bbb{P}_{i_{0}}\left( \tau _{0}=1\right) =\Bbb{P}%
_{i_{0}}\left( \tau _{0}>0\right) -\Bbb{P}_{i_{0}}\left( \tau _{0}>1\right) =%
\mathbf{e}_{i_{0}}^{\prime }\left( I-\overline{W}_{\beta }\right) \mathbf{1}%
=1-\overline{w}_{\beta }\left( i_{0}\right) ,$ the probability mass defect
of $\overline{W}_{\beta }$ at state $i_{0}.$

For all $\left( i_{0},i_{n}\right) $, we have: $\lim_{n\uparrow \infty
}\left[ \overline{W}_{\beta }^{n}\left( i_{0},i_{n}\right) \right] ^{1/n}=%
\overline{\rho }_{\beta }$ and, only when $A$ is primitive (irreducible and
aperiodic), by the strong version of Perron-Frobenius theorem (see \cite{HJ}%
) 
\begin{equation}
\lim_{n\uparrow \infty }\overline{\rho }_{\beta }^{-n}\overline{W}_{\beta
}^{n}=\mathbf{\phi }_{\beta }\mathbf{\pi }_{\beta }^{\prime },  \label{15b}
\end{equation}
where $\mathbf{\pi }_{\beta }^{\prime }>0$ and $\mathbf{\phi }_{\beta }>0,$
defined in (\ref{6}), are the left- (right- ) eigenvectors of $\overline{W}%
_{\beta }$ associated to $\overline{\rho }_{\beta }$, chosen, as before, so
as to satisfy $\mathbf{\pi }_{\beta }^{\prime }\mathbf{\phi }_{\beta }=1.$
As a result of (\ref{15a}) and (\ref{15b}), when $A$ is primitive: 
\begin{equation}
\lim_{n\uparrow \infty }\overline{\rho }_{\beta }^{-n}\Bbb{P}_{i_{0}}\left(
\tau _{0}>n\right) =\phi _{\beta }\left( i_{0}\right) ,  \label{16}
\end{equation}
meaning that $\tau _{0}$ is \emph{tail-equivalent to a geometric random
variable} \emph{with success probability} $\overline{\rho }_{\beta }$. The
latter formula therefore gives the limiting interpretation of $\mathbf{\phi }%
_{\beta }$ in the context of the RW $\left\{ X_{n}\right\} $. What about $%
\mathbf{\pi }_{\beta }$?\newline

Firstly, because $\mathbf{\pi }_{\beta }^{\prime }$ is the left
eigenprobability vector of $\overline{W}_{\beta }$ with eigenvalue $%
\overline{\rho }_{\beta }:$ 
\begin{equation}
\Bbb{P}_{\mathbf{\pi }_{\beta }}\left( \tau _{0}>n\right)
:=\sum_{i_{0}=1}^{N}\pi _{\beta }\left( i_{0}\right) \Bbb{P}_{i_{0}}\left(
\tau _{0}>n\right) =\mathbf{\pi }_{\beta }^{\prime }\overline{W}_{\beta }^{n}%
\mathbf{1}=\overline{\rho }_{\beta }^{n}.  \label{18}
\end{equation}
If the process is started with $\mathbf{\pi }_{\beta }$, the law of $\tau
_{0}$ is \emph{exactly} geometrically distributed on $\left\{
1,2,...,N\right\} $ with success probability $\overline{\rho }_{\beta }.$%
\newline

Consider now the conditional probability $\Bbb{P}_{i_{0}}\left(
X_{n}=i_{n}\mid \tau _{0}>n\right) $.

Recalling $\Bbb{P}_{i_{0}}\left( X_{n}=i_{n},\tau _{0}>n\right) =\mathbf{e}%
_{i_{0}}^{\prime }\overline{W}_{\beta }^{n}\mathbf{e}_{i_{n}},$ by Bayes
rule, we get 
\begin{equation*}
\Bbb{P}_{i_{0}}\left( X_{n}=i_{n}\mid \tau _{0}>n\right) =\frac{\mathbf{e}%
_{i_{0}}^{\prime }\overline{W}_{\beta }^{n}\mathbf{e}_{i_{n}}}{\mathbf{e}%
_{i_{0}}^{\prime }\overline{W}_{\beta }^{n}\mathbf{1}}=\frac{\mathbf{e}%
_{i_{0}}^{\prime }\left( \overline{\rho }_{\beta }^{-n}\overline{W}_{\beta
}^{n}\right) \mathbf{e}_{i_{n}}}{\mathbf{e}_{i_{0}}^{\prime }\left( 
\overline{\rho }_{\beta }^{-n}\overline{W}_{\beta }^{n}\right) \mathbf{1}},
\end{equation*}
showing that, independently of the starting point $i_{0}$ 
\begin{equation*}
\Bbb{P}_{i_{0}}\left( X_{n}=i_{n}\mid \tau _{0}>n\right) \underset{n\uparrow
\infty }{\rightarrow }\frac{\mathbf{e}_{i_{0}}^{\prime }\left( \mathbf{\phi }%
_{\beta }\mathbf{\pi }_{\beta }^{\prime }\right) \mathbf{e}_{i_{n}}}{\mathbf{%
e}_{i_{0}}^{\prime }\left( \mathbf{\phi }_{\beta }\mathbf{\pi }_{\beta
}^{\prime }\right) \mathbf{1}}=\pi _{\beta }\left( i_{n}\right) .
\end{equation*}
Such a probability measure $\mathbf{\pi }_{\beta }$ is called a Yaglom limit 
\cite{Y} of $\left\{ X_{n}\right\} $.

Further, with $\Bbb{P}_{\mathbf{\pi }_{\beta }}\left( \cdot \right)
:=\sum_{i=1}^{N}\pi _{\beta }\left( i_{0}\right) \Bbb{P}_{i_{0}}\left( \cdot
\right) $, for each $n$, $i_{n}\in \left\{ 1,2,..,N\right\} :$%
\begin{equation}
\Bbb{P}_{\mathbf{\pi }_{\beta }}\left( X_{n}=i_{n}\mid \tau _{0}>n\right) :=
\label{19}
\end{equation}
\begin{equation*}
\frac{\Bbb{P}_{\mathbf{\pi }_{\beta }}\left( X_{n}=i_{n},\tau _{0}>n\right) 
}{\Bbb{P}_{\mathbf{\pi }_{\beta }}\left( \tau _{0}>n\right) }=\frac{\mathbf{%
\pi }_{\beta }^{\prime }\overline{W}_{\beta }^{n}\mathbf{e}_{i_{n}}}{\mathbf{%
\pi }_{\beta }^{\prime }\overline{W}_{\beta }^{n}\mathbf{1}}=\pi _{\beta
}\left( i_{n}\right) ,
\end{equation*}
and this precisely means that $\mathbf{\pi }_{\beta }$ is the (unique)
quasi-stationary distribution (QSD) of $\left\{ X_{n}\right\} $. As is
well-known for Markov chains with finite state-space absorbed at $\partial $%
, we observe that the Yaglom limit coincides with its QSD. \emph{When }$A$%
\emph{\ is primitive (strongly connected and aperiodic), Equations (\ref{18}%
), (\ref{19}) and (\ref{16}) provide a natural interpretation of }$\rho
_{\beta },$\emph{\ }$\mathbf{\pi }_{\beta }$\emph{\ and }$\mathbf{\phi }%
_{\beta }$ \emph{in terms of the RW governed by }$P$\emph{\ in (\ref{15})
and its stopping time }$\tau _{0}.$

We refer to \cite{L} for additional informations on QSD and Yaglom limits in
the context of population dynamics.\newline

\textbf{Remark:} When $A$ is irreducible but not primitive (the underlying
topological graph is strongly connected but periodic), only the following
weaker form of the Perron-Frobenius theorem holds true \cite{HJ} 
\begin{equation*}
\lim_{K\uparrow \infty }\frac{1}{K}\sum_{n=1}^{K}\overline{\rho }_{\beta
}^{-n}\overline{W}_{\beta }^{n}=\mathbf{\phi }_{\beta }\mathbf{\pi }_{\beta
}^{\prime }.
\end{equation*}
With $\phi _{\beta }^{+}=\max_{i}\phi _{\beta }\left( i\right) $, $\phi
_{\beta }^{-}=\min_{i}\phi _{\beta }\left( i\right) $, Equation (\ref{16})
has to be weakened into 
\begin{equation*}
\phi _{\beta }^{-}/\phi _{\beta }^{+}\leq \overline{\rho }_{\beta }^{-n}\Bbb{%
P}_{i_{0}}\left( \tau _{0}>n\right) \leq \phi _{\beta }^{+}/\phi _{\beta
}^{-},\text{ for all }n\text{, }i_{0}.
\end{equation*}
The quantity $\overline{\rho }_{\beta }^{-n}\Bbb{P}_{i_{0}}\left( \tau
_{0}>n\right) $ may oscillate and not tend to some limit; however $-\frac{1}{%
n}\log \Bbb{P}_{i_{0}}\left( \tau _{0}>n\right) \rightarrow \overline{\rho }%
_{\beta }$ still holds true. $\vartriangle $\newline

The above construction of the RW $\left\{ X_{n}\right\} $ allows now to
interpret two fundamental $W_{\beta }-$consistent RWs.\emph{\newline
}

\textbf{The locally conditioned random walk:} With $D_{\overline{\mathbf{w}}%
_{\beta }}:=$ diag$\left( \overline{\mathbf{w}}_{\beta }\right) $, the
transition matrix of the one-step conditioned process $\left( X_{1}\mid \tau
_{0}>1\right) $ is: 
\begin{equation*}
\Pi \left( i,j\right) =\frac{\mathbf{e}_{i}^{\prime }\overline{W}_{\beta }%
\mathbf{e}_{j}}{\mathbf{e}_{i}^{\prime }\overline{W}_{\beta }\mathbf{1}}=%
\mathbf{e}_{i}^{\prime }\left[ D_{\overline{\mathbf{w}}_{\beta }}^{-1}%
\overline{W}_{\beta }\right] \mathbf{e}_{j}=\frac{\mathbf{e}_{i}^{\prime
}W_{\beta }\mathbf{e}_{j}}{\mathbf{e}_{i}^{\prime }W_{\beta }\mathbf{1}}=%
\mathbf{e}_{i}^{\prime }\left[ D_{\mathbf{w}_{\beta }}^{-1}W_{\beta }\right] 
\mathbf{e}_{j},
\end{equation*}
normalizing each line $i$ by $\mathbf{e}_{i}^{\prime }\overline{W}_{\beta }%
\mathbf{1}=\overline{w}_{\beta }\left( i\right) .$ Clearly, $\Pi \mathbf{1}=%
\mathbf{1}$ and the RW with transition matrix : 
\begin{equation}
\Pi =D_{\mathbf{w}_{\beta }}^{-1}W_{\beta }  \label{20}
\end{equation}
is $W_{\beta }-$consistent. Note that $\Pi $ is \emph{invariant} under the
scaling $W_{\beta }\rightarrow \overline{W}_{\beta }.$

Let $\mathbf{\mu }$ be the invariant associated to this $\Pi $. It holds, 
\cite{RO}, that 
\begin{equation*}
\mu \left( i\right) =\frac{\left( I-\Pi \right) _{i,i}}{\sum_{i}\left( I-\Pi
\right) _{i,i}}
\end{equation*}
where $\left( I-\Pi \right) _{i,i}$ is the cofactor of the $\left(
i,i\right) -$entry of the matrix $I-\Pi $. Then, if $\overleftarrow{\Pi }$
is the transition matrix of the reversed (backward in time) chain of $\Pi $
at equilibrium: $\overleftarrow{\Pi }^{\prime }=D_{\mathbf{\mu }}\Pi D_{%
\mathbf{\mu }}^{-1}.$ In general, $\overleftarrow{\Pi }\neq \Pi $ and
detailed balance may not hold. \newline

\textbf{Global conditioning and the canonical process. }Consider now the
proper Markov chain whose transition probabilities are obtained by the Doob
transform: 
\begin{equation*}
\Pi _{*}\left( i,j\right) =\overline{\rho }_{\beta }^{-1}\frac{\phi _{\beta
}\left( j\right) }{\phi _{\beta }\left( i\right) }\overline{W}_{\beta
}\left( i,j\right) =\rho _{\beta }^{-1}\frac{\phi _{\beta }\left( j\right) }{%
\phi _{\beta }\left( i\right) }W_{\beta }\left( i,j\right) ,\text{ }i,j\in
\left\{ 1,..,N\right\} ,
\end{equation*}
satisfying $\Pi _{*}\mathbf{1}=\mathbf{1}.$ In matrix form: 
\begin{equation}
\Pi _{*}=\rho _{\beta }^{-1}D_{\mathbf{\phi }_{\beta }}^{-1}W_{\beta }D_{%
\mathbf{\phi }_{\beta }},  \label{21}
\end{equation}
and $\Pi _{*}$ is also invariant under the scaling $W_{\beta }\rightarrow 
\overline{W}_{\beta }.$ An important property of this RW is the following:
The probability $\Pi _{*}\left( \mathbf{i}_{n}\right) :=\prod_{m=1}^{n}\Pi
_{*}\left( i_{m-1},i_{m}\right) $ of the $n-$path $\mathbf{i}_{n}$ is 
\begin{equation*}
\Pi _{*}\left( \mathbf{i}_{n}\right) =\rho _{\beta }^{-n}W_{\beta }\left( 
\mathbf{i}_{n}\right) \prod_{m=1}^{n}\frac{\phi _{\beta }\left( i_{m}\right) 
}{\phi _{\beta }\left( i_{m-1}\right) }=\rho _{\beta }^{-n}W_{\beta }\left( 
\mathbf{i}_{n}\right) \frac{\phi _{\beta }\left( i_{n}\right) }{\phi _{\beta
}\left( i_{0}\right) }.
\end{equation*}
For a bridge $n-$path for which $i_{0}=i_{n},$ $\Pi _{*}\left( \mathbf{i}%
_{n}\right) =\rho _{\beta }^{-n}W_{\beta }\left( \mathbf{i}_{n}\right) $
reduces, up to a scaling constant, to the weight $W_{\beta }\left( \mathbf{i}%
_{n}\right) $ of the $n-$path $\mathbf{i}_{n}.$

The invariant probability distribution $\mathbf{\mu }_{*}$ on $\left\{
1,..,N\right\} $ satisfying $\mathbf{\mu }_{*}^{\prime }\Pi _{*}=\mathbf{\mu 
}_{*}^{\prime }$ exists. It is given explicitly by $\mathbf{\mu }_{*}=%
\mathbf{\pi }_{\beta }*\mathbf{\phi }_{\beta }$ and so: 
\begin{equation}
\mu _{*}\left( i\right) =\pi _{\beta }\left( i\right) \phi _{\beta }\left(
i\right) \text{, }i=1,..,N.  \label{21a}
\end{equation}
Doob transforms have to do with conditioning a process on its lifetime. The
Markov chain with one-step transition probability matrix $\Pi _{*}$ may be
shown to be the one of the process whose one-step transition probability
distribution is: 
\begin{equation}
\Pi _{*}\left( i,j\right) =\lim_{n\uparrow \infty }\Bbb{P}_{i}\left(
X_{1}=j\mid \tau _{0}>n\right) ,  \label{22}
\end{equation}
corresponding to $X_{n}$ \emph{conditioned} to never hit the coffin state $%
\partial =\left\{ 0\right\} $ in the distant future; see \cite{L}. This
process has a unique invariant measure given by $\mathbf{\mu }_{*}$ in (\ref
{21a}).

Defining as before $\overleftarrow{\Pi }_{*}$ by: $\overleftarrow{\Pi }%
_{*}^{\prime }=D_{\mathbf{\mu }_{*}}\Pi _{*}D_{\mathbf{\mu }_{*}}^{-1}$, we
can ask conditions under which detailed balance $\overleftarrow{\Pi }_{*}=$ $%
\Pi _{*}$ holds. We have 
\begin{equation*}
\overleftarrow{\Pi }_{*}^{\prime }=\rho _{\beta }^{-1}D_{\mathbf{\pi }%
_{\beta }}D_{\mathbf{\phi }_{\beta }}D_{\mathbf{\phi }_{\beta
}}^{-1}W_{\beta }D_{\mathbf{\phi }_{\beta }}D_{\mathbf{\phi }_{\beta
}}^{-1}D_{\mathbf{\pi }_{\beta }}^{-1}=\rho _{\beta }^{-1}D_{\mathbf{\pi }%
_{\beta }}W_{\beta }D_{\mathbf{\pi }_{\beta }}^{-1}
\end{equation*}
so that $\overleftarrow{\Pi }_{*}=\rho _{\beta }^{-1}D_{\mathbf{\pi }_{\beta
}}^{-1}W_{\beta }^{\prime }D_{\mathbf{\pi }_{\beta }}$ showing that
reversibility holds when $W_{\beta }=W_{\beta }^{\prime }$ since if this is
the case: $\mathbf{\pi }_{\beta }=\mathbf{\phi }_{\beta }.$\newline

\textbf{No extra state. }We emphasize here that there are some alternative
ways to force the substochastic problem into a stochastic one. Assume $%
A\left( i,i\right) >0$ for each $i$, in which case $A^{N-1}>0$ and $A$
necessarily is primitive. Consider the stochastic matrix $\Pi $ which is $%
W_{\beta }-$consistent: 
\begin{equation}
\Pi =\overline{W}_{\beta }+D_{\mathbf{1-}\overline{\mathbf{w}}_{\beta }},
\label{23}
\end{equation}
where $D_{\mathbf{1-}\overline{\mathbf{w}}_{\beta }}:=$ diag$\left( \mathbf{1%
}-\overline{\mathbf{w}}_{\beta }\right) $, satisfying $\Pi \mathbf{1}=%
\mathbf{1}$. In that case, the mass defect vector $\mathbf{1-}\overline{%
\mathbf{w}}_{\beta }$ is transferred to the diagonal entries of $\overline{W}%
_{\beta }$ to make it stochastic, without appealing to an extra
coffin-state. Note that $\Pi $ in (\ref{23}) \emph{no} longer is \emph{%
invariant} under the scaling $W_{\beta }\rightarrow \overline{W}_{\beta }$
and so this normalization is norm-dependent$.$

\section{Special cases}

\textbf{1.} \textbf{The topological case.} Assume $\beta =0$. In this case, $%
W_{0}=A$ and 
\begin{equation*}
\log \rho _{0}>s=-\sum_{i}\mu \left( i\right) \sum_{j}\Pi \left( i,j\right)
\log \Pi \left( i,j\right)
\end{equation*}
for all $A-$consistent matrix $\Pi \neq \Pi _{*}^{0}$ with: $\Pi
_{*}^{0}\left( i,j\right) =\frac{1}{\rho _{0}}A\left( i,j\right) \frac{\phi
_{0}\left( j\right) }{\phi _{0}\left( i\right) }.$ $\log \rho _{0}>0$
interprets as the maximal entropy production rate of all Markov chains
governed by such $\Pi $s. The RW with transition matrix $\Pi _{*}^{0}$ is
termed the maximal entropy random walk in \cite{BDLW}. Its invariant measure
is $\mu _{*}\left( i\right) =\psi _{0}\left( i\right) ^{2}.$ When $\Pi =D_{%
\mathbf{a}}^{-1}A,$ with $\mathbf{a}=A\mathbf{1}$, the invariant measure is $%
\mu \left( i\right) =a\left( i\right) /\sum_{i}a\left( i\right) ,$
proportional to the node degrees. Then $s=\sum_{i}a\left( i\right) \log
a\left( i\right) /\sum_{i}a\left( i\right) $ and $\log \rho _{0}>s$ is an
inequality first discussed in \cite{BDLW}. When disorder is present, the
canonical RW associated to $W_{0}=A$ was also shown therein to exhibit \emph{%
localization} properties.

Consider the general inequality (\ref{11a}) for all $W_{\beta }-$consistent
stochastic matrix $\Pi \neq \Pi _{*}=:$ $\frac{1}{\rho _{\beta }}W_{\beta
}\left( i,j\right) \frac{\phi _{\beta }\left( j\right) }{\phi _{\beta
}\left( i\right) }.$ Choosing for $\Pi $ the above particular value: $\Pi
=\Pi _{*}^{0},$ we obtain 
\begin{equation*}
\log \rho _{\beta }>\log \rho _{0}-\beta \sum_{i}\psi _{0}\left( i\right)
^{2}\sum_{j}\Pi _{*}^{0}\left( i,j\right) H\left( i,j\right) .
\end{equation*}
\emph{Therefore, the average transition energy } 
\begin{equation}
\alpha _{0}:=\sum_{i}\psi _{0}\left( i\right) ^{2}\sum_{j}\Pi _{*}^{0}\left(
i,j\right) H\left( i,j\right)  \label{24}
\end{equation}
\emph{interprets as the slope at }$\beta =0$\emph{\ of the graph pressure
function }$\beta \rightarrow $\emph{\ }$p\left( \beta \right) $, namely: $%
\alpha _{0}=p^{^{\prime }}\left( 0\right) .$ We have $f\left( \alpha
_{0}\right) =-p\left( 0\right) =\log \rho _{0}.$\newline

\textbf{2.} \textbf{The potential case}: Assume $H\left( i,j\right) =U\left(
j\right) -U\left( i\right) $ for some potential $U$ attached to the nodes of
the graph. In this case, the matrix $K_{\beta }$ defining $W_{\beta }$ is
called a potential kernel. Firstly, in this case, it follows from (\ref{24})
and the equilibrium property of $\left( \mu _{*}=\psi _{0}*\psi _{0},\Pi
_{*}^{0}\right) $ that: 
\begin{equation}
\alpha _{0}:=\sum_{i}\psi _{0}\left( i\right) ^{2}\sum_{j}\Pi _{*}^{0}\left(
i,j\right) \left[ U\left( j\right) -U\left( i\right) \right] =0.  \label{25}
\end{equation}
We conclude that \emph{the singularity spectrum of all graph with potential
kernel }$K_{\beta }$\emph{\ attains its maximum at} $\alpha _{0}=0$.\newline

With $\mathbf{v}_{\beta }$ the column-vector with entries $v_{\beta }\left(
i\right) =\exp -\beta U\left( i\right) $, we get: 
\begin{equation}
W_{\beta }:=A*K_{\beta }=D_{\mathbf{v}_{\beta }}^{-1}AD_{\mathbf{v}_{\beta
}}.  \label{26}
\end{equation}
We have $\mathbf{w}_{\beta }=W_{\beta }\mathbf{1}=D_{\mathbf{v}_{\beta
}}^{-1}A\mathbf{v}_{\beta }$ and $D_{\mathbf{w}_{\beta }}=D_{\mathbf{v}%
_{\beta }}^{-1}D_{A\mathbf{v}_{\beta }}.$ Note that $W_{\beta }$ is
diagonally similar to $A$ so that the spectral radius of $W_{\beta }$ is $%
\rho _{0}$, independently of $\beta .$\newline

- Consider first the RW with transition matrix $\Pi =D_{\mathbf{w}_{\beta
}}^{-1}W_{\beta }=D_{A\mathbf{v}_{\beta }}^{-1}AD_{\mathbf{v}_{\beta }}$.
Its invariant measure is characterized by: $\mathbf{\mu }^{\prime }=\mathbf{%
\mu }^{\prime }\Pi $. Recalling $A=A^{\prime }$, we find $\mathbf{\mu }%
\propto D_{\mathbf{v}_{\beta }}A\mathbf{v}_{\beta },$ with normalized
entries weighting output degree nodes with lowest $U$: 
\begin{equation}
\mu \left( i\right) =\sum_{j}A\left( i,j\right) e^{-\beta \left(
U_{i}+U_{j}\right) }/\sum_{i,j}A\left( i,j\right) e^{-\beta \left(
U_{i}+U_{j}\right) }.  \label{27}
\end{equation}
This RW with transition matrix $\Pi $ is reversible because $\overleftarrow{%
\Pi }=D_{\mathbf{\mu }}^{-1}\Pi ^{\prime }D_{\mathbf{\mu }}=\Pi .$\newline

- Secondly, consider the canonical RW consistent with $W_{\beta }=D_{\mathbf{%
v}_{\beta }}^{-1}AD_{\mathbf{v}_{\beta }}$. The right eigenvector $\mathbf{%
\phi }_{\beta }$ of $W_{\beta }=D_{\mathbf{v}_{\beta }}^{-1}AD_{\mathbf{v}%
_{\beta }}$ is $\mathbf{\phi }_{\beta }=D_{\mathbf{v}_{\beta }}^{-1}\mathbf{%
\phi }_{0}.$ It is associated to the eigenvalue $\rho _{0}$. Thus the
canonical RW has transition matrix $\Pi _{*}$ is given by: 
\begin{equation}
\Pi _{*}=\rho _{0}^{-1}D_{\mathbf{\phi }_{\beta }}^{-1}W_{\beta }D_{\mathbf{%
\phi }_{\beta }}=\rho _{0}^{-1}D_{\mathbf{\phi }_{0}}^{-1}AD_{\mathbf{\phi }%
_{0}}=\Pi _{*}^{0}.  \label{28}
\end{equation}
Its invariant measure is $\mathbf{\pi }_{*}=\mathbf{\psi }_{0}*\mathbf{\psi }%
_{0}.$ \emph{The canonical RW consistent with the potential weight matrix }$%
W_{\beta }=D_{\mathbf{v}_{\beta }}^{-1}AD_{\mathbf{v}_{\beta }}$\emph{\
always coincides with the canonical RW consistent with its adjacency matrix }%
$A$ \emph{governed by }$\Pi _{*}^{0}.$\newline

With $\Pi =D_{\mathbf{w}_{\beta }}^{-1}W_{\beta }$ with entries 
\begin{equation*}
\Pi \left( i,j\right) =A\left( i,j\right) e^{-\beta \left( U\left( j\right)
-U\left( i\right) \right) }/\sum_{j}A\left( i,j\right) e^{-\beta \left(
U\left( j\right) -U\left( i\right) \right) },
\end{equation*}
and with invariant measure $\mu \left( i\right) $ displayed in (\ref{27}),
for all $\beta ,$ we get 
\begin{eqnarray*}
\log \rho _{0} &>&-\sum_{i}\mu \left( i\right) \sum_{j}\Pi \left( i,j\right)
\log \Pi \left( i,j\right) -\beta \sum_{i}\mu \left( i\right) \sum_{j}\Pi
\left( i,j\right) \left( U\left( j\right) -U\left( i\right) \right) \\
&=&-\sum_{i}\mu \left( i\right) \sum_{j}\Pi \left( i,j\right) \log \Pi
\left( i,j\right) =s.
\end{eqnarray*}
\emph{We conclude that for potential kernels }$K_{\beta }$\emph{, the
entropy production rate of the RW with probability transition matrix }$\Pi
=D_{\mathbf{w}_{\beta }}^{-1}W_{\beta }=D_{\mathbf{w}_{\beta }}^{-1}\left[
A*K_{\beta }\right] $\emph{\ is always bounded above by }$\log \rho _{0}.$%
\newline

\textbf{Remark:} The $W_{\beta }-$consistent RW with transition matrix $\Pi
=D_{\mathbf{w}_{\beta }}^{-1}W_{\beta }$ associated to the weight kernel $%
W_{\beta }=AD_{\mathbf{v}_{\beta }}$ was also considered in \cite{GL}. For
this model, the cost of a transition from $i$ to $j$ only depends on the
terminal state, regardless of where one starts from. Although the latter is
not in the potential class, its invariant measure is also given by (\ref{27}%
). $\triangledown $\newline

\textbf{3.} \textbf{The symmetric case}. If $H\left( i,j\right) =H\left(
j,i\right) $, then $W_{\beta }=W_{\beta }^{\prime }$ itself$.$ For example $%
H\left( i,j\right) =\left| U\left( j\right) -U\left( i\right) \right| $ for
some potential $U$ attached to the nodes of the graph, or $H\left(
i,j\right) $ is some distance (ultrametric or not) between nodes $i$ and $j$%
. In this case, for all $\beta $, the invariant measure $\mathbf{\mu }_{*}$
of the canonical RW governed by $\Pi _{*}=\frac{1}{\rho _{\beta }}D_{\mathbf{%
\phi }_{\beta }}^{-1}W_{\beta }D_{\mathbf{\phi }_{\beta }}$, is: 
\begin{equation}
\mathbf{\mu }_{*}=\mathbf{\psi }_{\beta }*\mathbf{\psi }_{\beta }  \label{29}
\end{equation}
and the corresponding RW is reversible.

When $W_{\beta }=W_{\beta }^{\prime }$, the invariant measure associated to $%
\Pi =D_{\mathbf{w}_{\beta }}^{-1}W_{\beta }$ satisfying $\mathbf{\mu }%
^{\prime }=\mathbf{\mu }^{\prime }\Pi $ is given by: 
\begin{equation*}
\mu \left( i\right) =\frac{w_{\beta }\left( i\right) }{\sum_{i}w_{\beta
}\left( i\right) }=\frac{\sum_{j}A\left( i,j\right) e^{-\beta H\left(
i,j\right) }}{\sum_{i,j}A\left( i,j\right) e^{-\beta H\left( i,j\right) }}.
\end{equation*}
We have $\overleftarrow{\Pi }^{\prime }=D_{\mathbf{\mu }}\Pi D_{\mathbf{\mu }%
}^{-1}=D_{\mathbf{w}_{\beta }}D_{\mathbf{w}_{\beta }}^{-1}W_{\beta }D_{%
\mathbf{w}_{\beta }}^{-1}=W_{\beta }D_{\mathbf{w}_{\beta }}^{-1}=\Pi
^{\prime }$ so that $\overleftarrow{\Pi }=\Pi :$ detailed balance also holds.


\begin{thebibliography}{99}
\bibitem{AF}  Aldous, D.; Fill, J. Reversible Markov Chains and Random Walks
on Graphs. book in preparation, 2000.

\bibitem{AGD}  Arnold, L.; Gundlach, V.M.; Demetrius, L. Evolutionary
formalism for products of positive random matrices. The Annals of Applied
Probability, Vol. 4, No 3, 859-901, (1994).

\bibitem{BDLW}  Burda, Z.; Duda, J.; Luck, J.M.; Waclaw, B. Localization of
maximal entropy random walk. arXiv:0810.4113 (2008).

\bibitem{FK}  Friedland, S.; Karlin, S. Some inequalities for the spectral
radius of non-negative matrices and applications. Duke Math. J. 42, no. 3,
459--490, (1975).

\bibitem{GL}  Gomez-Gardenes, J.; Latora, V. Entropy rate of diffusion
processes on complex networks. Physical Review E, 78, 065102(R), (2008).

\bibitem{HJ}  Horn, R.A.; Johnson, C.R. Matrix analysis. Cambridge
University Press, Cambridge, 1985.

\bibitem{L}  Lambert, A. Population dynamics and random genealogies.
Stochastic Models 24, suppl. 1, 45--163, (2008).

\bibitem{RAC}  Rached, Z.; Alajaji, F.; Campbell, L. R\'{e}nyi's divergence
and entropy rates for finite alphabet Markov sources. IEEE Transactions on
Information Theory 47(4): 1553-1561 (2001).

\bibitem{RO}  Romanovsky, V. I. Discrete Markov chains. Translated from the
Russian by E. Seneta, Wolters-Noordhoff Publishing, Groningen 1970.

\bibitem{R}  Ruelle, D. Thermodynamic formalism. The mathematical structures
of equilibrium statistical mechanics. Second edition. Cambridge Mathematical
Library. Cambridge University Press, Cambridge, 2004.

\bibitem{Y}  Yaglom, A. M. Certain limit theorems of the theory of branching
random processes. Doklady Akad. Nauk SSSR (N.S.) 56, 795--798, (1947).
\end{thebibliography}
\end{document}